# Ambient and high pressure phases of tin sulfide compounds

Joseph M. Gonzalez, Kien Nguyen-Cong, Brad A. Steele, and Ivan I. Oleynik

*Department of Physics, University of South Florida, Tampa, FL 33620*

(Dated: September 18, 2017)

Tin sulfides, $Sn_xS_y$, are important class of materials that are actively investigated as novel photovoltaic and water splitting materials. First-principles evolutionary crystal structure search is performed with the goal of constructing the complete phase diagram of $Sn_xS_y$ and discover new phases as well as new compounds of varying stoichiometry at ambient conditions and pressures up to 100 GPa. The ambient phase of $SnS_2$ with $P\bar{3}m1$ symmetry remains stable up to 30 GPa. Another ambient phase, SnS experiences a series of phase transformations: including $\alpha$-SnS to $\beta$-SnS at 9 GPa, followed by $\beta$-SnS to $\gamma$-SnS at 40 GPa. $\gamma$-SnS is a new high-pressure metallic phase with $Pm\bar{3}m$ space group symmetry stable up to 100 GPa, which becomes a superconductor with a maximum $T_c$ =9.6 K at 40 GPa. Another new metallic compound, $Sn_3S_4$ with $I\bar{4}3d$ space group symmetry, is predicted to be stable at pressures above 15 GPa, which also becomes a superconductor with a maximum $T_c$ =21.8 K at 30 GPa.

## I. INTRODUCTION

Tin sulfur binary compounds, $Sn_xS_y$, represent an emergent class of electronic materials that have been studied for several decades[1-7]. Three compounds are known to exist at ambient conditions, $\alpha$-SnS, $SnS_2$ and $Sn_2S_3$, all of which are semiconducting materials[8]. Of particular interest is $\alpha$-SnS, which adopts a layered orthorhombic structure with $Pnma$ symmetry at ambient conditions. Similar to other group IV-VI metal chalcogen materials, SnS is semiconducting with an indirect band gap of $\sim 1$ eV and relatively strong optical absorption[9] at photon energies greater than 1.3 eV. Due to its exceptional electronic and vibrational properties, $\alpha$-SnS is actively investigated for use in novel electronic[10], photovoltaic[11,12] and thermoelectric[13-17] applications.

Upon increasing the pressure and/or temperature $\alpha$-SnS-$Pnma$ is known to experience a structural transformation to $\beta$-SnS which is semi-metallic phase with $Cmcm$ space group symmetry. While the transition temperature of the $\alpha$-SnS $\rightarrow$ $\beta$-SnS transformation is well established, $\sim 900$ K[18], the transition pressure is not, with predictions of 15 GPa[19], in disagreement with experimental value of 10.5 GPa[10]. In addition to $\alpha$ and $\beta$ phases, SnS is reported to exist in several other phases such as rock-salt structure with space group symmetry $Fm\bar{3}m$[9] as well as the recently reported $\pi$-SnS phase with large cubic unit cell and space group symmetry $P2_13$[20].

Another important material is $SnS_2$, which adopts a layered structure with $P\bar{3}m1$ symmetry at ambient conditions and exists in both bulk and single-layer form[5,21]. $SnS_2$-$P\bar{3}m1$ is unique compared to $\alpha$-SnS in that several polytypes exist, i.e. 2H, 4H, and 3R, which have different stacking sequences that arise depending on the growth conditions[4]. $SnS_2$-$P\bar{3}m1$ is a well known semiconductor with an indirect band gap of $\sim 2$ eV[21], which has received significant attention due to the unique properties arising from layered structure of $SnS_2$, including relatively high carrier mobility[5], gas sensing[6], exceptionally large excitonic effects[21]. Another unique layer dependent effect seen in $SnS_2$ is a decrease in thermal conductivity with simultaneous increase in the electrical conductivity with decreasing layer count[22].

The orthorhombic $Sn_2S_3$ crystal with $Pnma$ symmetry, is another interesting semiconducting compound which possesses a band gap of $\sim 1$ eV and relatively a large optical absorption[8] at ambient conditions. In contrast to SnS and $SnS_2$, $Sn_2S_3$ does not have a layered structure although it shares structural units from $SnS_2$-$P\bar{3}m1$ crystal.

While the $Sn_xS_y$ family of materials has been extensively studied at ambient and relatively low pressures, their high pressure behavior has been relatively unexplored. High pressure synthesis is currently pursued as an effective method to activate unusual chemistry, resulting in novel compounds and phases that are difficult to synthesize using traditional methods at ambient conditions[23-26]. In addition to helping synthesis of new materials, the properties of known materials can be dramatically altered under high pressure, as demonstrated by the discovery of high temperature superconductivity in sulfur hydride[27] and metallicity of hydrogen[28]. Therefore, the goal of this work is to explore the entire compositional phase space of the $Sn_xS_y$ system as a function of pressure. In particular, we seek to discover new compounds and phases both at ambient conditions and high pressure up to 100 GPa. The electronic properties of predicted compounds including superconducting behavior of few metallic phases are also investigated.

## II. COMPUTATIONAL DETAILS

The search for ambient and high pressure $Sn_xS_y$ compounds with variable stoichiometry is conducted using the first-principles evolutionary structure search code USPEX[29,30]. The USPEX method has been demonstrated to be highly effective in predicting many novel phases in a wide range of materials under ambient and high pressure conditions[23,25,26,31,32]. To properly sample the compositional phase space of $Sn_xS_y$, variable-composition search is performed. At each pressure, the first generation of individuals is produced by randomly sampling the stoichiometry, space group, lattice parameters and atomic positions after which the energy of the structures is minimized using standard density functional theory (DFT). Future generations of structures are constructed from a small percentage of random structures as well as by applying variation operators, i.e. atom mutation, lattice mutation, heredity, etc., to the lowest enthalpy individuals from the previous generation. The generations are produced and optimized until there is no change in the lowest enthalpy structures over several generations, thus resulting in structures that are presumably the most energetically favorable.

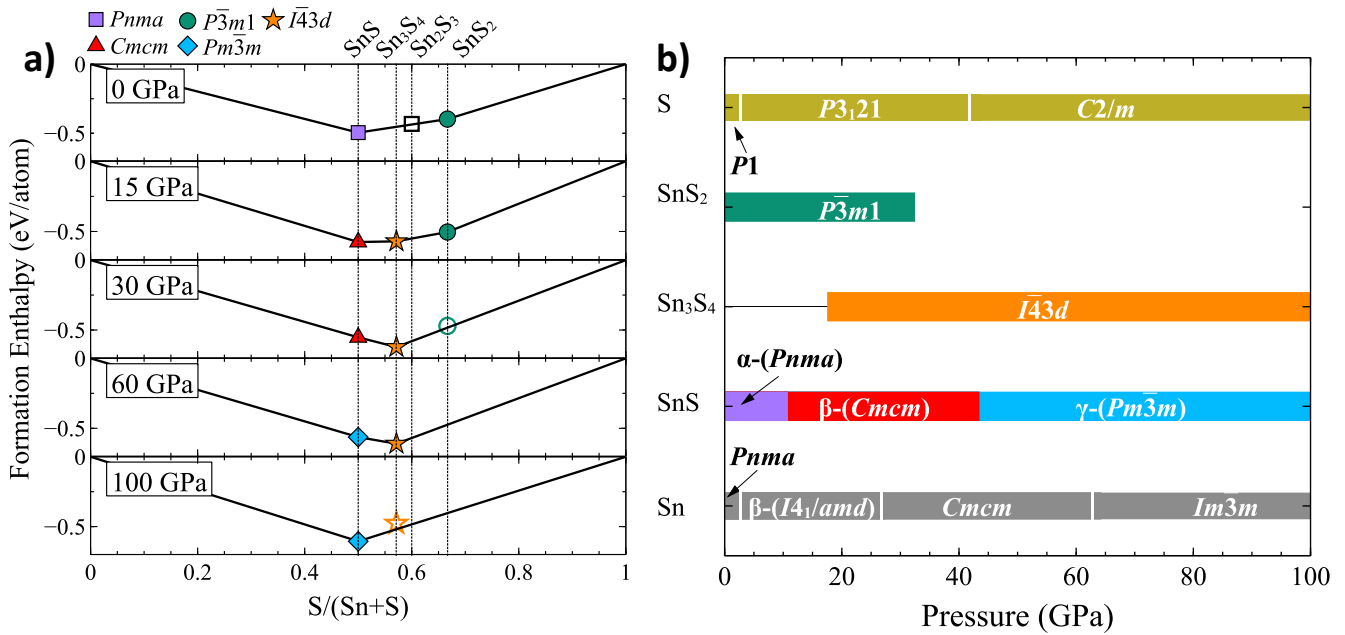

FIG. 1. a) Convex hulls at 0, 15, 30, 60 and 100 GPa. For all pressures, the open symbols denote metastable structures close to the convex hull. b) Composition-pressure phase diagram of the $Sn_xS_y$ system up to 100 GPa.

To quantify the thermodynamic stability of the predicted structures at a given pressure, the formation enthalpy is calculated, which is the difference of enthalpy of each compound and the sum of enthalpies of the lowest enthalpy phases of pure elements at corresponding pressure. These elemental phases are $Pnma$, $I4/amd$, $Cmcm$ and $Im\bar{3}m$ for tin, and $P1$, $P3_21$, and $C2/m$ for sulphur depending on the pressure. Within the set of $Sn_xS_y$ compounds, those with the lowest formation enthalpy define the convex hull - the formation enthalpy/composition curve, see the curves in Figure 1(a). The series of convex hulls corresponding to several pressures are used to construct the $Sn_xS_y$ phase diagram, shown in Figure 1(b).

First principles DFT calculations are performed using the Vienna *ab initio* simulation package (VASP)[33] using the projector augmented wave (PAW) pseudopotentials, and Perdew, Burke, and Ernzenhof (PBE) generalized gradient approximation (GGA) to DFT[34]. The weak van der Waals (vdW) interactions that play an impor-tant role in layered metal-chalcogenide compounds, are accounted for by employing the empirical Grimme D2 method[35]. During the search, the plane-wave cutoff is set to 450 eV and the sampling of the Brillouin zone is done with a $k$-point density of 0.06 Å$^{-1}$ to achieve the convergence of total energy to within 10 meV/atom. Once the search is complete, the enthalpy of candidate structures is then recalculated with a more accurate set of parameters: the plane-wave cutoff is set to 900 eV and k-point sampling density of 0.02 Å$^{-1}$ which provides energies within 5 meV/atom and a maximum force on any atom less than 0.03 eV/Å. Finally, dynamical stability is quantified by calculating the vibration spectra using the frozen-phonon technique as implemented in the Phonopy code[36].

## III. RESULTS

### A. Convex Hulls

The reliability of the USPEX method of structure prediction is validated by finding all known ground state compounds at ambient conditions, $\alpha$-SnS-$Pnma$ and $SnS_2$-$P\bar{3}m1$, without any prior input, see Figure 1(a). Two other compounds, $Sn_2S_3$-$Pnma$ and $\pi$-SnS, shown as open symbols in Figure 1(a), are not on the convex hull at 0 GPa, but above only by a few meV/atom, which indicates they are metastable. Although they are known to exist at ambient conditions[8,37], they were not found during the search due the limitation on maximum number of atoms (20) in a unit cell used in the search.

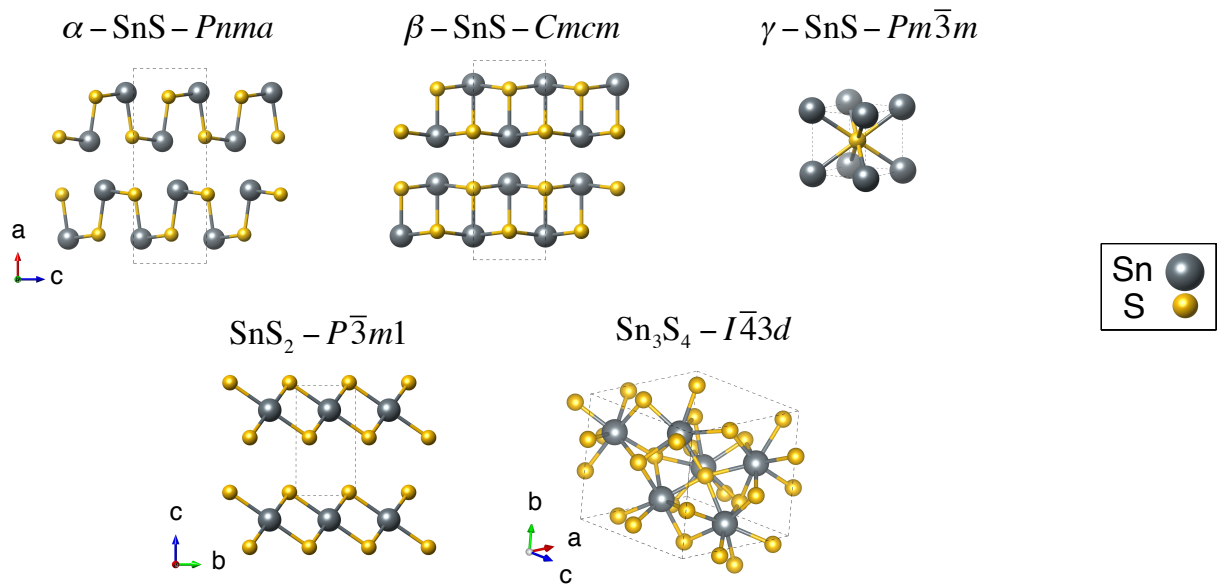

FIG. 2. Crystal structures of the lowest enthalpy compounds found during the search. The tin atoms are displayed as grey spheres and the sulfur atoms as yellow spheres.

In addition to ambient conditions, thermodynamically stable $Sn_xS_y$ compounds are searched at several pressure up to 100 GPa. According to the calculated convex hull at 15 GPa in Figure 1(a) $\alpha$-SnS-$Pnma$ is no longer the lowest enthalpy phase and transforms to $\beta$-SnS-$Cmcm$, see their crystal structures in Figure 2. In addition, the $SnS_2$-$P\bar{3}m1$ compound remains thermodynamically stable at 15 GPa. Another feature present on the 15 GPa convex hull is the emergence of a new compound with an unique stoichiometry, $Sn_3S_4$ with symmetry $I\bar{4}3d$, see Figure 2. This compound has been reported in experiment[38] but has not been thoroughly characterized. Increasing the pressure to 30 GPa, $SnS_2$-$P\bar{3}m1$ becomes thermodynamically unstable, i.e. is no longer on the convex hull, while the new compound, $Sn_3S_4$-$I\bar{4}3d$, becomes the lowest enthalpy structure at 30 GPa.

At 60 GPa, $\beta$-SnS-$Cmcm$ is no longer the lowest enthalpy phase and transforms to a new phase, $\gamma$-SnS, with $Pm\bar{3}m$ symmetry, see Figure 2. The $Sn_3S_4$-$I\bar{4}3d$ compound remains thermodynamically stable and is the lowest enthalpy structure at 60 GPa. Finally at 100 GPa, which is the highest pressure considered in this work, $\gamma$-SnS and appears to be the only thermodynamically stable compound. The crystal structures of all major compounds and phases discovered during the structure search are shown in Figure 2.

The composition-pressure phase diagram for $Sn_xS_y$ is shown in Figure 1(b). For each stoichiometry, the lowest enthalpy phase which is both thermodynamically and dynamically stable (i.e. does not have imaginary frequency phonon modes) is determined by calculating the phonon dispersion relation over the pressure range of thermodynamic stability. For example, the phase $SnS_2$-$P\bar{3}m1$ is determined to be stable from 0 to 28 GPa. Also shown in Figure 1(b) is the pressure range of stability for the newly predicted compound, $Sn_3S_4$-$I\bar{4}3d$ which is found to be stable between 18 and 100 GPa. The details of the transformation between phases of the SnS compound with the 1:1 stoichiometry upon increase of pressure up to 100 GPa is discussed separately in the following subsection.

## B. SnS phases

Figure 3(a) shows the calculated enthalpy differences up to 100 GPa for the different phases of SnS, relative to $\alpha$-SnS-$Pnma$ phase. At low pressures up to $\sim 9$ GPa, $\alpha$-SnS-$Pnma$ is the lowest enthalpy phase. However, the other phases are higher in enthalpy by only a few meV. For pressures greater than 10 GPa, $\beta$-SnS-$Cmcm$ remains the energetically preferred phase up to $\sim 39$ GPa. At this pressure a first-order phase transition to $\gamma$-SnS-$Pm\bar{3}m$ phase occurs with a dramatic reduction in volume per atom, see Figure 3(b). Another phase, SnS-$Fm\bar{3}m$, which is metastable at ambient conditions, and higher in enthalpy by $\sim 50$ meV/atom than the ground state phase $\alpha$-SnS-$Pnma$ has been synthesized experimentally[3] at 0 GPa. This small enthalpy difference is increased with pressure over the entire pressure range up to 100 GPa, see Figure 3.

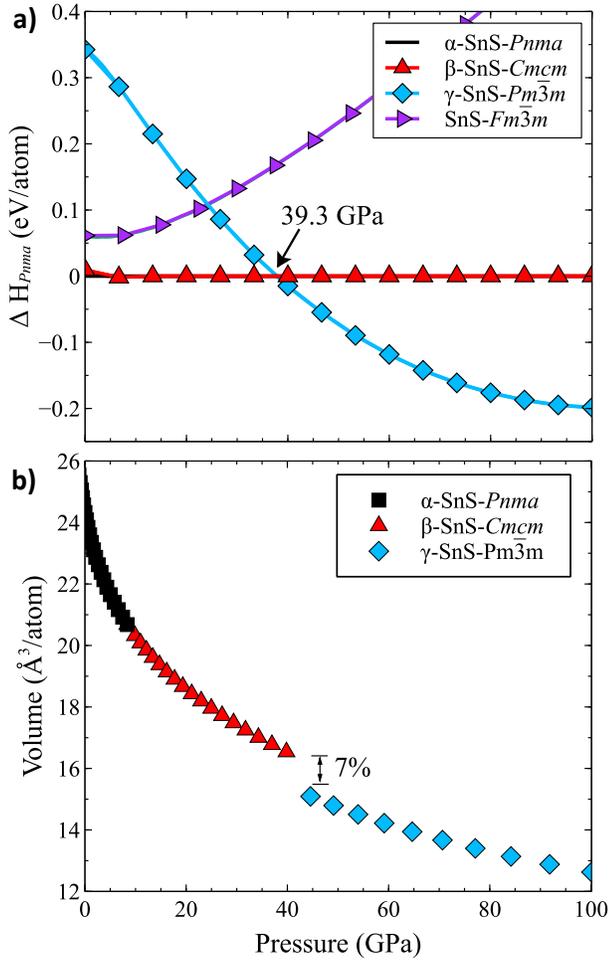

FIG. 3. a) Calculated enthalpy differences versus pressure for various phases of SnS referenced by the enthalpy of the ground state structure α-SnS-*Pnma*. b) Equation of state for the various phases of SnS.

As mentioned previously, there exists a discrepancy concerning the transition pressure at which the semiconducting α-SnS-*Pnma* phase transforms to semi-metallic β-SnS-*Cmcm* phase[10,19]. The latter phase is sometimes labeled as that having *Bbmm* space group symmetry (space group No. 63), which is a non-standard definition of *Cmcm* in which the lattice vectors possess the same ordering as in the *Pnma* phase, i.e. $b < c < a$. To determine unambiguously the phase transition pressure, several order parameters are defined to map the structural similarity between α-SnS-*Pnma* and β-SnS-*Cmcm*, see Figure 2(b). In particular, the trajectory of a tin atom can be monitored during the compression, shown in Figure 4(a). Between 0 GPa and 8.56 GPa, the fractional $z$-coordinate of the tin atom rapidly decreases from ∼ 0.1 to 0, at this pressure the position of the tin atom in the α-phase matches the position of the tin atom in the β-phase. Another useful order parameter characterizing the transition is the bond angle a tin atom makes with two sulfur atoms as a function of hydrostatic compression. At 0 GPa, the (S-Sn-S) bond angle in α-SnS-

*Pnma* is 90°, see Figure 4(c), which rapidly converges to the (S-Sn-S) bond angle in β-SnS-*Cmcm* at ∼ 9 GPa.

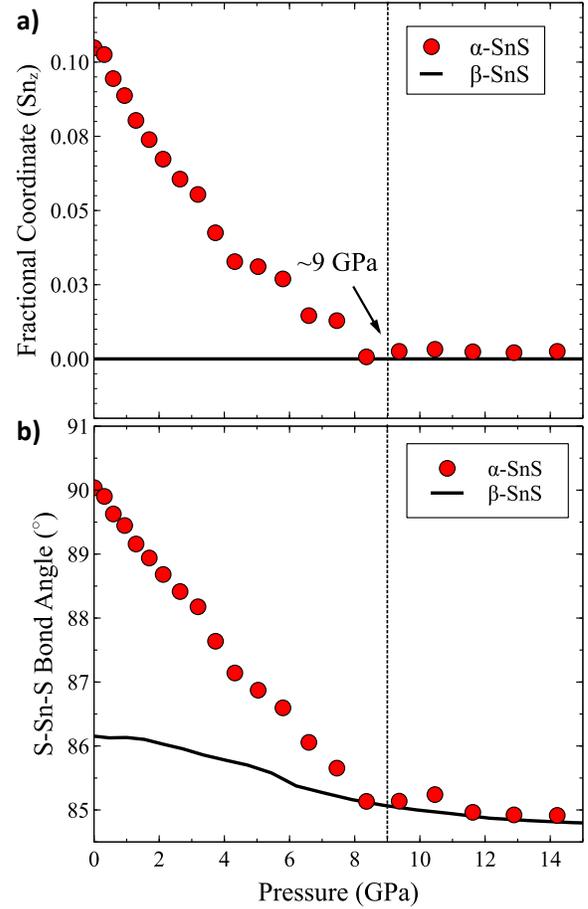

FIG. 4. Order parameters quantifying the transition α-SnS-*Pnma* → β-SnS-*Cmcm*. a) Fractional $z$-coordinate of an equivalent tin atom and b) the bond angle an equivalent tin atom makes with two sulfur atoms as a function of hydrostatic compression and c) crystal structure of α-SnS-*Pnma* and β-SnS-*Cmcm* clearly showing the bond angle being measured during compression.

As XRD is routinely used to monitor phase transformations at high pressures, the calculated XRD spectrum of α-SnS-*Pnma* as a function of pressure up to 9 GPa is presented in Figure 5. The evolution of the XRD pattern with increasing pressure shows a clear decrease of intensity of the (120) and (101) peaks, thus loosing the major features attributed to $\alpha$-SnS-$Pnma$, see Figure 5. In contrast to the case of SnSe[40], there are no experimental reports on the XRD measurements detailing $\alpha$-SnS-$Pnma$ to $\beta$-SnS-$Cmcm$ phase transitions, therefore, the calculated XRD patterns will be useful for guiding future experiments.

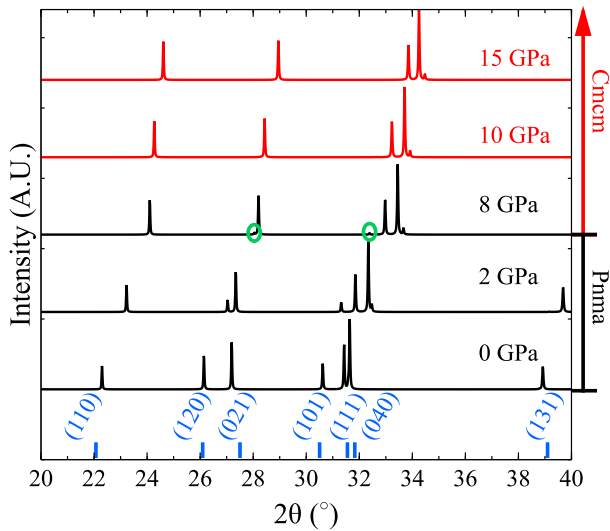

FIG. 5. Calculated X-Ray diffraction spectrum of SnS under hydrostatic compression up to 15 GPa. Green circles indicate the presence of the (120) and (101) peaks found in $\alpha$-SnS-$Pnma$. Shown in blue is the standard XRD pattern for $\alpha$-SnS-$Pnma$ , JCPDS No. 00-039-0354[39].

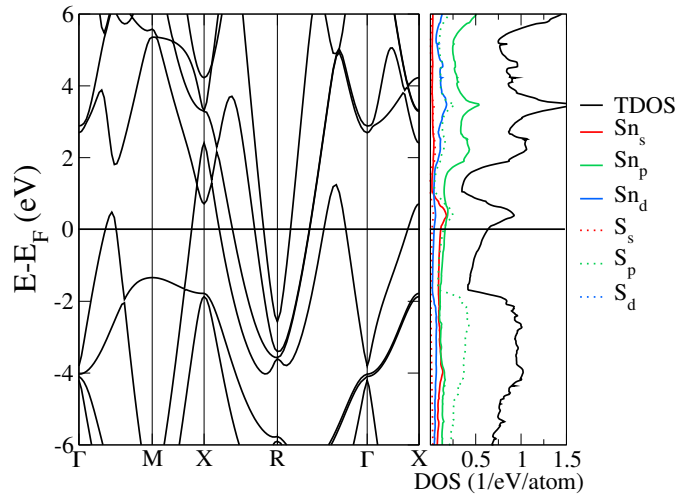

FIG. 6. Electronic band structure and projected density of states for $\gamma$-SnS-$Pm\bar{3}m$ at 40 GPa.

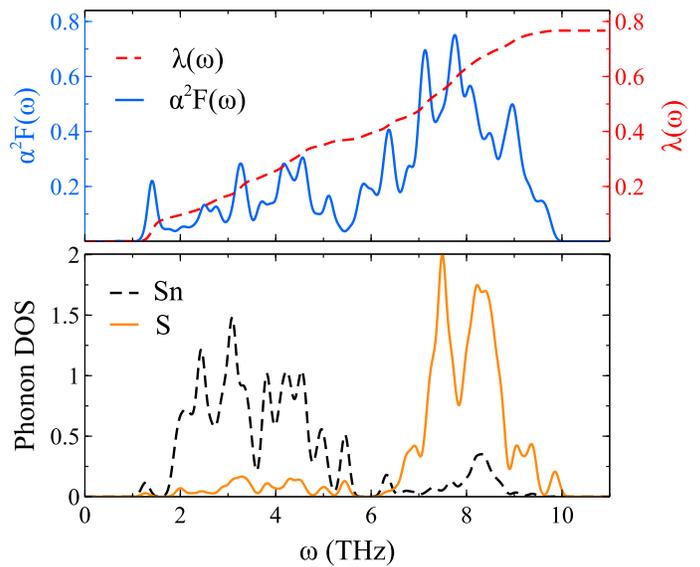

FIG. 7. a) Eliashberg spectral function $\alpha^2F(\omega)$ and integrated electron-phonon coupling strength $\lambda(\omega)$; and b) atomically resolved phonon density of states for $\gamma$-SnS-$Pm\bar{3}m$ at 40 GPa.

In addition to structural changes, the transformation from $\alpha$-SnS-$Pnma$ to $\beta$-SnS-$Cmcm$ is accompanied by changes in the electronic structure. At ambient conditions, bulk $\alpha$-SnS-$Pnma$ crystal is a semiconductor with a calculated indirect band gap of $E_g = 1.28$ eV, which decreases monotonically in magnitude with increasing pressure. During the compression, $\alpha$-SnS-$Pnma$ remains an indirect band gap semiconductor with $E_g$ changing relatively quickly upon increase of pressure, a pressure derivative being $\partial E_g/\partial P = -0.181$ eV/GPa. Using the pressure derivative and the ambient pressure band gap, $\alpha$-SnS-$Pnma$ is predicted to becomes a semi-metal at 6.8 GPa.

Figure 6 depicts the electronic band structure along with the electronic density of states projected onto atomic orbitals (PDOS) for $\gamma$-SnS-$Pm\bar{3}m$ phase at 40 GPa. $\gamma$-SnS displays a relatively large density of states at the Fermi level which implies it is metallic. Within the energy window between $-6$ and 6 eV around the Fermi level, the tin and sulfur atoms contribute equally to the electronic states in SnS, with a sharp peak just above the Fermi level. Also seen in Figure 6, the electronic states above the Fermi level mainly arise from the Sn $5p$ orbitals with a small contribution from the S $4p$ and Sn $4d$ orbitals. The main contribution to the electronic states below the Fermi level comes from the S $4p$ orbitals.

Until now, $\gamma$-SnS phase has not been predicted theo-retically or observed in experiment. However this phase is known to exist within the tin-selenide family at pressures greater than 50 GPa and is found to exhibit superconductivity with critical temperature $T_c = 5$ K at 58 GPa[41,42]. In the case of materials exhibiting standard Bardeen-Cooper-Schrieffer (BCS) superconductivity, its emergence arises due to coupling between the electronic and vibrational degrees of freedom, which is characterized by the Eliashberg electron-phonon spectral function, $\alpha^2F(\omega)$, see Figure 6(b). From the $\alpha^2F(\omega)$ dependence on phonon frequency $\omega$, many important microscopic quantities of BCS theory can be derived, namely the electron phonon coupling constant (EPC) $\lambda$, and the averaged logarithmic phonon frequency $\omega_{\log}$, which allows for the estimation of the critical temperature using the Allen-Dynes modified McMillan equation[43],

$$T_c = \frac{\omega_{\log}}{1.2} \exp\left[-\frac{1.04(1+\lambda)}{\lambda - \mu^*(1 + 0.62\lambda)}\right] \qquad (1)$$

Within this model, $\mu^*$ is the Coulomb repulsion parameter, which is usually within the range $0.1 - 0.2$ for normal metals. In the case of $\gamma$-SnS-$Pm\bar{3}m$ at 40 GPa, $\lambda$ and $\omega_{\log}$ are calculated to be 0.77 and 156.4 cm$^{-1}$(4.69 THz) respectively, see Supporting Information for details of these calculations. By using typical values of $\mu^* = 0.1$, $\gamma$-SnS-$Pm\bar{3}m$ is predicted to have a critical temperature of 9.6 K at 40 GPa. Upon increasing the pressure further to 100 GPa, the superconducting critical temperature drops dramatically to less than 1 K. Although the phonon frequencies increase with pressure, the coupling to the electronic structure is significantly diminished which results in a significant drop in T$_c$.

To understand the mechanism behind the superconducting state in $\gamma$-SnS-$Pm\bar{3}m$ at 40 GPa, the calculated Eliashberg spectral function and atomically resolved phonon density of states are calculated, see Figure 7. The spectral function for $\gamma$-SnS-$Pm\bar{3}m$ has a relatively large band centered just below 8 THz with another small band centered at 3 THz. Also shown in Figure 7(b) is the integrated electron-phonon coupling constant, $\lambda$, which is a measure of the strength of $\alpha^2F(\omega)$. To understand the main contributions to $\lambda$, the atomically resolved phonon density of states are presented in Figure 7(b). The frequency spectrum below 6 THz is dominated by vibrations due to the Sn atoms, while for higher frequencies, $6 < \omega < 10$ THz , the spectrum is dominated by S atoms. In particular, the Sn atoms contribute 51 % to $\lambda$ while S contributes 49 %.

## C. Sn$_3$S$_4$: A new metallic compound

Tin normally has two preferred oxidation states, +2 or +4, while sulfur typically has an oxidation state of $-2$. That is why these two elements preferably combine in 1:1 and 1:2 stoichiometries to form SnS and SnS$_2$ compounds at ambient conditions. However, under high pressure, the distribution of electrons is altered due to pressure-induced confinement, resulting in altered valences of Sn and S atoms. Indeed, the convex hull at 15 GPa, Figure 1(a), displays a new compound with a novel stoichiometry, Sn$_3$S$_4$-$I\bar{4}3d$, which is thermodynamically stable, see the conventional unit cell in Figure 2. To our knowledge, this is the first report of this unique material with 3:4 stoichiometry within the family of tin-sulfur compounds, although in case of selenium, Sn$_3$Se$_4$-$I\bar{4}3d$ has been observed[44]. Based on the phonon dispersion at different pressures Sn$_3$S$_4$-$I\bar{4}3d$ is dynamically unstable at 0 GPa, but with increasing the hydrostatic compression becomes dynamically stable at 15 GPa, see phonon dispersion in Supporting Information, Figure S1.

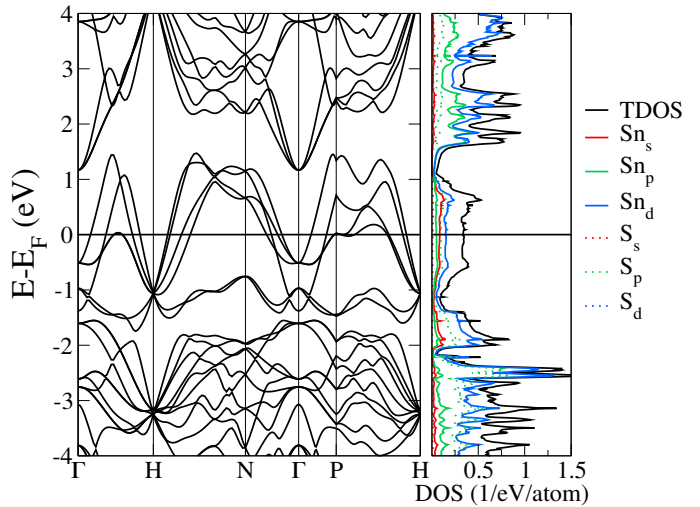

FIG. 8. Electronic band structure and projected density of states for Sn$_3$S$_4$-$I\bar{4}3d$ at 30 GPa.

The electronic structure of the newly predicted compound Sn$_3$S$_4$-$I\bar{4}3d$ presented in Figure 8 clearly displays its metallic character with a relatively large DOS at the Fermi level. In contrast to $\gamma$-SnS-$Pm\bar{3}m$, the electronic states of Sn$_3$S$_4$-$I\bar{4}3d$ above the Fermi level are populated by roughly equal contributions from the Sn $4d$ and $5p$ orbitals, whereas below the Fermi level, by equal amounts of the Sn $4d$ and S $4p$ orbitals.

Given the relatively large density of states near the Fermi level, Sn$_3$S$_4$-$I\bar{4}3d$ is also investigated as a poten-tial superconductor. The calculated Eliashberg spectral function is presented in Figure 9(a): in contrast to $\gamma$-SnS, there are three well defined peaks in the spectrum. In particular, there is one small band centered at 3 THz while there are two large but narrow bands centered at 7 THz and 9.5 THz. By integrating $\alpha^2 F(\omega)$, the strength of the electron-phonon coupling $\lambda$ is calculated to be 1.2 while the logarithmic frequency $\omega_{\log}$, is calculated to be 169.6 cm$^{-1}$ (5.1 THz). Similar to the case of $\gamma$-SnS-$Pm\bar{3}m$, $T_c$ is calculated with Eq. (1) using $\mu^* = 0.1$, and Sn$_3$S$_4$-$I\bar{4}3d$ is predicted to have a critical temperature of 21.8 K, see Supporting Information for details of these calculations. In contrast to $\gamma$-SnS-$Pm\bar{3}m$, the superconducting state has an increasing trend with increasing pressure; i.e. $T_c = 12$ K at 15 GPa and $T_c = 21.8$ K at 30 GPa, which results from an increase in phonon frequencies as well as the coupling between electronic and vibrational degrees of freedom.

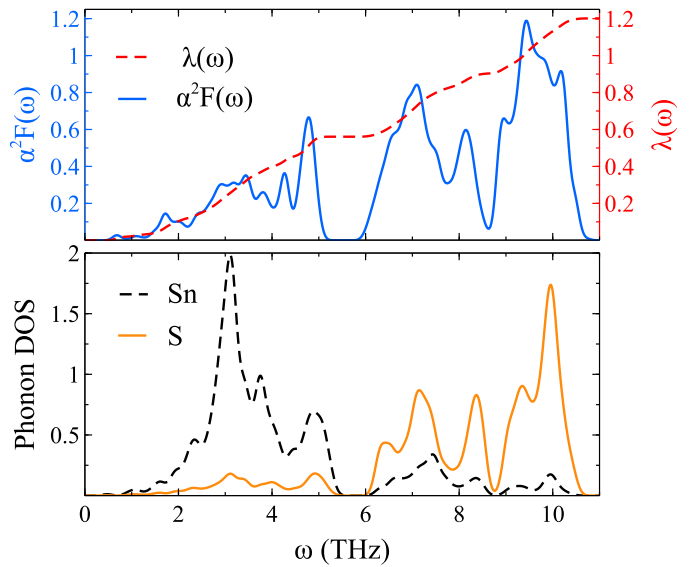

FIG. 9. a) Eliashberg spectral function $\alpha^2F(\omega)$ and integrated electron-phonon coupling constant $\lambda(\omega)$; b) phonon density of states for Sn$_3$S$_4$-$I\bar{4}3d$ at 30 GPa.

To understand the microscopic mechanism behind the superconductivity in Sn$_3$S$_4$-$I\bar{4}3d$ phase, the atomically resolved phonon densities of state are presented in Figure 9(b). Similar to $\gamma$-SnS-$Pm\bar{3}m$ phase, the frequency range below 6 THz is dominated by Sn atoms, while the higher frequencies are dominated by S atom. In contrast to $\gamma$-SnS-$Pm\bar{3}m$, the contributions to $\lambda$ are not equal with Sn contributing approximately 18 % while S contributing approximately 82 %. This result suggests that S plays a more significant role in superconductivity of Sn$_3$S$_4$-$I\bar{4}3d$ than in $\gamma$-SnS-$Pm\bar{3}m$. This is supported by the fact that although the frequency ranges are approximately equal, the stronger electron phonon coupling due to S atoms produces a higher $T_c$.

## IV. CONCLUSION

In summary, Sn$_x$S$_y$ compounds are systematically investigated at ambient conditions and under hydrostatic compression at pressures from 0 to 100 GPa. In case of 1:1 stoichiometry, three unique phases of SnS compounds are found to be stable: they undergo the following sequence of phase transitions $\alpha - $SnS$ - Pnma \rightarrow \beta$ at 9 GPa and $\beta - $SnS$ - Cmcm \rightarrow \gamma - $SnS$ - Pm\bar{3}m$ at 39.3 GPa. For pressures higher than 40 GPa, a new phase $\gamma$-SnS-$Pm\bar{3}m$ is predicted. This new high pressure metallic phase is thermodynamically and dynamically stable up to the highest pressure studied, 100 GPa. In addition, $\gamma$-SnS-$Pm\bar{3}m$ is predicted to be a superconductor with T$_c$ =9.64 K at 40 GPa.

A new compound with 3:4 stoichiometry, Sn$_3$S$_4$-$I\bar{4}3d$, is predicted to become thermodynamically and dynamically stable from 15 GPa up to the highest pressure studied, 100 GPa. It is metallic and superconducting with a maximum T$_c$ =21.8 K at 30 GPa, which are a lower pressure and a higher temperature than those for $\gamma$-SnS. The prediction of stable novel compounds and phases at high pressures within the Sn$_x$S$_y$ family of binary compounds demonstrates the need for exploration of unex-pected compounds and phases of other metal-chalcogen materials.

We have also completed a comprehensive study of the pressure-composition phase diagram for the tin-selenide (Sn$_x$Se$_y$) family of materials and found that for most of the compounds, their crystal structures and stoichiometries are similar to those found in this work. This is not unexpected as S and Se are isoelectronic elements. However the ranges of pressure in the composition-pressure phase diagram are different. The quantitative information included in these phase diagrams provides the key information to experimentalists in future exploration of these materials.

## ACKNOWLEDGMENTS

The research is supported by the Defense Threat Reduction Agency, (grant No. HDTRA1-12- 1-0023) and Army Research Laboratory through Cooperative Agreement W911NF-16-2-0022. Simulations were performed using the NSF XSEDE supercomputers (grant No. TG-MCA08X040), DOE BNL CFN computational user facility, 9 and USF Research Computing Cluster sup- ported by NSF (grant No. CHE-1531590).

## APPENDIX A: SUPERCONDUCTING PROPERTIES

Superconducting properties were calculated using the Quantum ESPRESSO package. The plane-wave cutoff is set to 60 Ry and norm conserving pseudopotentials are used. For accurate sampling of the Brillouin zone, a 12x12x12 mesh is used, while a coarse q-point grid of 6x6x6 is used to calculate the phonon dynamical matrix elements. To estimate $T_c$, we use the Allen-Dynes modified McMillan formula

$$T_c = \frac{\omega_{\log}}{1.2} \exp\left[-\frac{1.04(1+\lambda)}{\lambda - \mu^*(1+0.62\lambda)}\right], \qquad (2)$$

where $\lambda$ is the electron-phonon coupling constant, $\omega_{\log}$ is the average logarithmic phonon frequency $\omega_{\log}$. In this formalism, $\mu^*$ is the Coulomb repulsion parameter, which has typical values between 0.1-0.14. To calculate $\lambda$, the following expression is used:

$$\lambda = \int d\omega \frac{\alpha^2 F(\omega)}{\omega},$$

where $\alpha^2 F(\omega)$ is Eliashberg spectral function. Finally, $\omega_{\log}$ is calculated as follows,

$$\omega_{\log} = \exp\left[\frac{2}{\lambda} \int d\omega \frac{\alpha^2 F(\omega)}{\omega} \ln(\omega)\right].$$

### APPENDIX B: DYNAMICAL STABILITY OF
### SN$_3$S$_4$-$I\bar{4}3d$

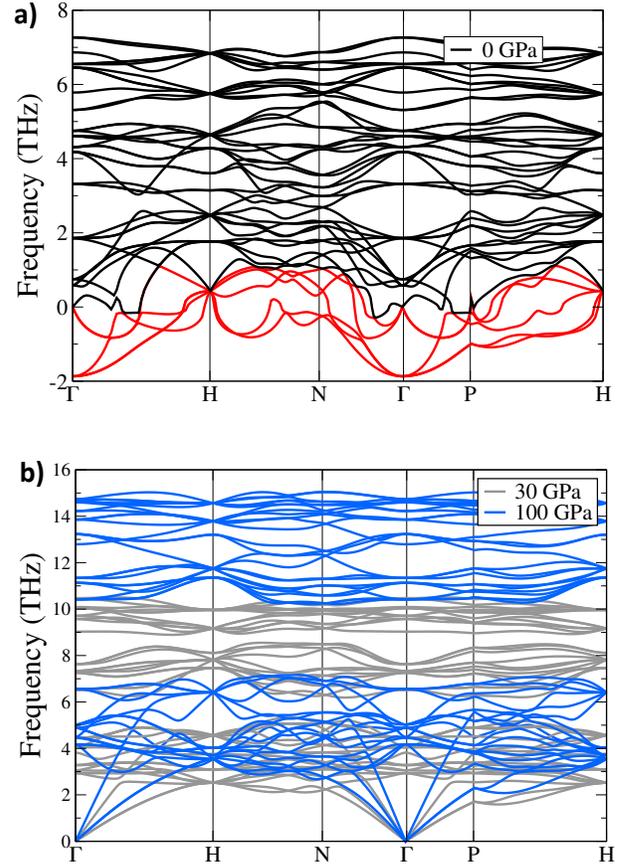

FIG. 10. Phonon dispersion for the new metallic compound Sn$_3$S$_4$-$I\bar{4}3d$ at a) 0 GPa and b) 30 GPa and 100 GPa. Red lines in 0 GPa phonon dispersion display imaginary phonon frequencies at ambient pressure.

### APPENDIX C: STRUCTURAL PARAMETERS
### OF TIN-SULFIDE COMPOUNDS

Table I. The structural parameters for each of the unique phases and compounds, which are stable within the pressure range of 0 to 100 GPa.

| Compound | Lattice Parameters (Å$^3$) | Atomic Positions |
|---|---|---|
| $\alpha$-SnS-$Pnma$ @0 GPa | $a = 11.39$ $b = 4.01$ $c = 4.26$ $\alpha = \beta = \gamma = 90°$ | Sn (4c) 0.126 0.250 -0.395 S (4c) 0.353 0.250 -0.481 |
| $\beta$-SnS-$Cmcm$ @10 GPa | $a = 3.89$ $b = 10.75$ $c = 3.85$ $\alpha = \beta = \gamma = 90°$ | Sn (4c) 0.000 0.120 0.250 S (4c) 0.000 0.357 0.250 |
| $\gamma$-SnS-$Pm\bar{3}m$ @40 GPa | $a = 3.126$ $b = 3.126$ $c = 3.126$ $\alpha = \beta = \gamma = 90°$ | Sn (1b) 0.500 0.500 0.500 S (1a) 0.000 0.000 0.000 |